\documentclass[preprint,12pt]{elsarticle}

\usepackage{amssymb}
\usepackage{amsmath}
\usepackage{wasysym}
\usepackage{psfrag}
\usepackage{color}
\usepackage[normalem]{ulem}

\newcommand{\lgr}{\ell_g}
\newcommand{\upd}{\mathrm{d}}
\newcommand{\dl}{\mbox{$\Delta \ell$}}
\newcommand{\dL}{\mbox{$\Delta L$}}
\newcommand{\dLmin}{\mbox{$\Delta L$}_{\mathrm{min}}}
\newcommand{\linf}{\ell_{\infty}}
\newcommand{\Linf}{L_{\infty}}
\newcommand{\Lcrit}{L_{c}}
\newcommand{\Ltot}{L_{\mathrm{tot}}}
\newcommand{\rhos}{\rho_s}
\newcommand{\beq}{\begin{equation}}
\newcommand{\eeq}{\end{equation}}

\journal{Extreme Mechanics Letters}

\begin{document}

\begin{frontmatter}



\title{The role of extensibility in the birth of a ruck in a rug}

\author{Alpha A. Lee, Cl\'{e}ment Le Gouellec and Dominic Vella}

\address{Mathematical Institute, Andrew Wiles Building, University of Oxford\\ Woodstock Rd, Oxford, OX2 6GG, UK}

\begin{abstract}
Everyday experience suggests that a `ruck' forms when the two ends of a heavy carpet or rug are brought closer together. Classical analysis, however, shows that the horizontal compressive force needed to create such a ruck should be infinite. We show that this apparent paradox is due to the assumption of inextensibility of the rug. By accounting for a finite extensibility, we show that rucks appear with a finite, non-zero end-shortening and confirm our theoretical results with simple experiments. Finally, we note that the appropriate measure of extensibility, the stretchability, is in this case not determined purely by geometry, but incorporates the mechanics of the sheet.
\end{abstract}


\end{frontmatter}


\section{Introduction}

Localized bumps (or rucks) in a carpet are an everyday annoyance but have also been used as an analogy to understand a plethora of physical phenomena involving the sliding of two solid bodies. Examples include dislocations in plastic deformation \cite{Nabarro1967}, Schallamach waves in rubber friction \cite{Schallamach1971,Briggs1975,Comninou1978}, and even slip pulses in earthquakes \cite{Hough2002}. Recently, a number of authors have focussed on the motion of such rucks, which can occur either rapidly (as when a rug is quickly shaken at one end) \cite{Vella2009,OKeefe2015} or slowly (as when a ruck `falls' down an inclined plane) \cite{Kolinski2009,Balmforth2015}. Surprisingly, however, some features of a static ruck remain poorly understood, including the conditions under which they form.

At a superficial level, the formation, or birth, of a ruck in a rug is similar to the buckling of the classic Euler elastica \cite{Levien2008}: the two ends of a rug are brought closer together by a distance $\dl$ and a ruck forms to accommodate the resulting excess length. Unlike the Euler elastica, however, the amount of  buckled material is not equal to the whole system size --- some material remains in contact with the `floor' and the arc-length of the ruck, $l$, is not known \emph{a priori}.

Everyday experience suggests that $l$ depends on the end--end compression $\dl$. To determine this relationship at the scaling level we follow Kolinski \emph{et al.} \cite{Kolinski2009}. For small rucks, simple geometry suggests that the height  of the ruck $d\sim (l\dl)^{1/2}$. For a rug of density $\rhos$, thickness $t$ and bending rigidity $B$, we expect that the gravitational energy of the ruck $\sim \rhos g t\times d\times l$, while the bending energy $\sim B(d/l^2)^2\times l$. Balancing these energies and eliminating $d\sim (l\dl)^{1/2}$ we find that
\beq
\dl\sim \left(\frac{\rhos g t}{B}\right)^2l^7.
\label{eqn:dl_inextens}
\eeq As expected, the width of the ruck  grows with increasing end--end compression.

How much compressive \emph{force} is required for the onset of rucking? A simple calculation \cite{Vella2009} reveals that the compressive force $T$ required to form a ruck of size $l$ satisfies
\begin{equation}
\frac{T l^2}{B} \approx80.76,
\label{classical_buckling}
\end{equation}  with $B$ the bending rigidity. Note that \eqref{classical_buckling} is precisely the classical result for the Euler buckling of a rod of length $l$ with clamped ends \cite{landau}.  Using \eqref{eqn:dl_inextens} to eliminate $l$ from \eqref{classical_buckling} in favor of $\dl$ reveals that the buckling load
\beq
T\sim \frac{B^{3/7}(\rhos gt)^{4/7}}{\dl^{2/7}},
\label{eqn:infiniteload}
\eeq which is divergent in the limit of very small end-shortenings $\dl$. To form a ruck from a flat rug we must pass through arbitrarily small end-shortenings and so we are led to the paradoxical result that to do so requires an infinite compressive force!

This divergence in the compressive force needed to form a ruck has been known for almost thirty years \cite{Wang1986}. Previous authors have attempted to explain it as a result of a breakdown of the linearized beam theory used to obtain \eqref{eqn:dl_inextens} and \eqref{classical_buckling} \cite{Kerr1973,ElAini1975,Kerr1978} while \cite{Wang1981} confirmed the validity of the linearized approach and showed decisively that a critical buckling load does not exist for a perfect, infinite, continuous horizontal heavy elastica. Here, we show using a combination of theory and experiment that the paradox is resolved by incorporating the finite extensibility of the material. 

A key assumption in the preceding discussion is that the object, the rug in this case, has a fixed arc length, i.e.~that it is inextensible. While this is true for objects with arbitrarily small thickness, any real object has a finite, if small thickness, and so is, to a certain extent, extensible. The effect of finite extensibility on the buckling of the classical Elastica has been studied by a number of authors \cite{Coffin1999,Neukirch2012,Pandey2014}. These analyses reveal that the crucial parameter governing the importance of extensibility is the ratio of the thickness, $t$, and total length, $\Ltot$, of the beam or, equivalently, the von K\'{a}rm\'{a}n number, $\gamma=t^2/(12\Ltot^2)$. However, the effect of $\gamma$ on the buckling threshold and the post-buckling behaviour is, in fact fairly small \cite{Neukirch2012}. An unusual exception is the vibration frequencies of some modes of buckled beams, which may be different for perfectly inextensible beams ($\gamma=0$) and asymptotically inextensible beams ($\gamma\to0$): the limit is singular \cite{Neukirch2012}.

In this context, the classic prediction that the compressive force  required to give  birth to a ruck diverges clearly suggests the possibility that  the finite extensibility must ultimately play a role in regularizing the problem; we expect that inclusion of this effect will lead to a finite compressive force  and a finite ruck size at the onset of buckling. Nevertheless, this suggestion appears not to have been made previously \cite{Wang1986} and so we  study this problem here. Using a combination of theory and tabletop experiments, we show that indeed the finite extensibility of any physical (rather than idealized) heavy elastica does give rise to a finite ruck height at birth and, hence, a finite buckling load. This is similar to the problem of the growth of a rod inside a curved cylinder where it was recently shown that finite extensibility regularizes the otherwise infinite pressure that would be applied on the cylinder prior to buckling \cite{Napoli2015}. However, unlike this related problem, and previous studies of the role of extensibility, we find that the relevant parameter governing the role of extensibility is \emph{not} simply determined by the geometrical properties of the object (namely $t$ and $\Ltot$) but depends also on its material properties (namely its Young's modulus $E$ and density $\rhos$).

\section{Theoretical analysis}

\subsection{Problem formulation}

\begin{figure}
\centering
\includegraphics[width=12cm]{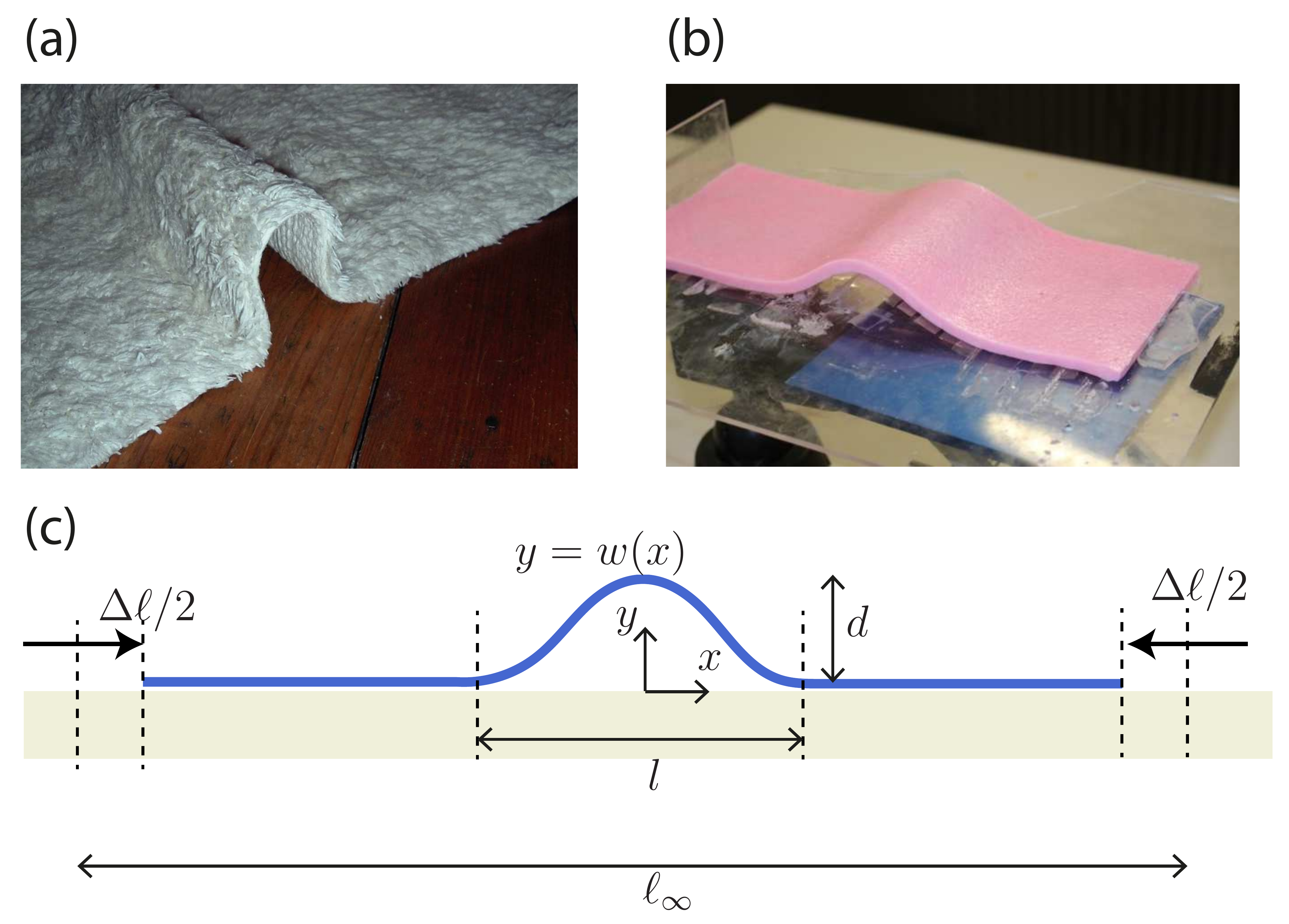}
\caption{(a) A rug placed on a floor often ends up with a bump (a ruck) in it caused by the relative motion of its ends. (b) A laboratory ruck formed using sheets of Polyvinylsiloxane. For scale, the horizontal dimension of the square blue plate here is $15\mathrm{~cm}$. (c) Schematic sketch of the model problem considered here. A heavy sheet of natural length $\linf$, resting on a horizontal, frictionless surface, forms a ruck of lateral size $l$ and height $d$ when its are displaced by a relative distance $\dl$.}
\label{fig:schematic}
\end{figure} 

We model the rug as a heavy elastic sheet of solid density $\rhos$, length $\linf$, thickness $t$ and width $b$. This sheet rests on a solid horizontal surface, which we assume to be frictionless for simplicity, and  is compressed by bringing the two ends a horizontal distance $\dl$ closer together (see figure~\ref{fig:schematic}c). The sheet can respond to this compression in two ways: by compressing along its length and by buckling out of the plane. If the sheet buckles, it does so over a finite region of width $l<\linf$ because the weight per unit area of the sheet ($\rhos g t$) opposes the whole sheet losing contact with the surface. Assuming small  transverse displacements, the profile of the sheet $y=w(x)$ satisfies the linearized heavy elastica  equation \cite{Vella2009,Wang1986}
\begin{equation}
B \frac{\upd^4w}{\upd x^4} + T \frac{\upd^2w}{\upd x^2} = -\rho g t
\label{beam} 
\end{equation}  for $|x|<l/2$. For $l/2\leq|x|<\linf/2$ we have $w=0$: the sheet is in contact with the substrate. 

In \eqref{beam}, $B = E t^3/[12(1-\nu^2)]$ is the bending stiffness of the sheet per unit width, with $\nu$  the Poisson Ratio. As the problem is symmetric about $x=0$, we consider only $0\leq x <\linf/2$, for simplicity. This gives rise to the symmetry boundary conditions
\begin{equation}
w'(0) = w'''(0) = 0. 
\label{sym_BC}
\end{equation} 
Since $w=0$ for $l/2<|x|<\linf/2$, continuity of vertical displacement, together with force and torque balance at $x=l/2$ then give 
\begin{equation}
w(l/2) = w'(l/2) = w''(l/2) = 0.
\label{contact_BC} 
\end{equation}

The system \eqref{beam}--\eqref{contact_BC} is ostensibly a fourth order system with five boundary conditions. However, the size of the buckled region, $l$, is unknown as is the imposed compressive force $T$ (since we are assuming that the sheet is subject to a known displacement of its ends). In fact, then, we have a sixth order problem with the final condition coming from a relationship between compression and tension. Ordinarily, one might be tempted to specify that the imposed compression should be accounted for by the out of plane displacement, so that $\dl=\int_0^{l/2}(w')^2~\upd x$. However, this condition implicitly assumes that the sheet is inextensible. Instead we return to Hooke's law, which relates $T$ to the strain, $e_{xx}$ within the sheet
\begin{equation}
T = - \frac{E t}{1-\nu^2} e_{xx}.
\label{hooke}
\end{equation} 
 Since the strain can be expressed in terms of the horizontal and out-of-plane displacements $u$ and $w$, respectively, we then have
\begin{equation}
 - \frac{1-\nu^2}{E t}T=e_{xx} = \frac{\mathrm{d} u}{\mathrm{d} x} + \frac{1}{2} \left(  \frac{\upd w}{\upd x}\right)^{2}.
\end{equation}  
Integrating from $0$ to $\linf/2$, we find that 
\begin{equation}
- \frac{\linf (1-\nu^2)}{2 E t}  T = \frac{1}{2}  \int_{0}^{\linf/2}\; \left(  \frac{\mathrm{d} w}{\mathrm{d} x}\right)^{2} \mathrm{d}x - \frac{\Delta l}{2},
\label{hooke_stress}
\end{equation} where we have used $u(\linf/2)=-\dl/2$ as the imposed displacement. (In the above derivation we have used that the compressive force $T$ is constant; Hooke's law \eqref{hooke} then shows that the strain $e_{xx}$ is also constant.)

The ordinary differential equation \eqref{beam} together with boundary conditions \eqref{sym_BC}--\eqref{contact_BC} and the global constraint \eqref{hooke_stress} completely specify the problem. To make further progress, however, we first non-dimensionalize the problem using the elasto-gravity length
\begin{equation}
\lgr = \left( \frac{B}{\rhos g t}\right)^{1/3}
\end{equation} to rescale all lengths; $\lgr$ emerges from the balance between bending and gravity that is expressed in \eqref{beam}. We  also introduce  the dimensionless compressive force $\sigma \equiv T \lgr^2/ B$.

The dimensionless equivalent of  \eqref{beam} in the non-contact (buckled) region is
\begin{equation}
W'''' + \sigma W'' = - 1.
\label{beam_nondim}
\end{equation} The boundary conditions \eqref{sym_BC}--\eqref{contact_BC} are unchanged, while \eqref{hooke_stress} becomes
\begin{equation}
- S \sigma \Linf =  \int_{0}^{\Linf/2}\; \left(  \frac{\mathrm{d} W}{\mathrm{d} X}\right)^{2} \mathrm{d}X - \dL, 
\label{strain_stretch}
\end{equation}
where $\{X,W,L,\Linf\}\equiv \{x,w,l,\linf\}/\lgr$ and
\begin{equation}
S \equiv \frac{t^2}{12 \lgr^2}, 
\label{eqn:Stretchability}
\end{equation} is the dimensionless `stretchability' of the sheet.  Note that while \eqref{eqn:Stretchability} has the form of the von-K\'{a}rm\'{a}n number ($\gamma=t^2/12l^2$)  discussed in the Introduction, the relevant horizontal length scale in the definition of $S$ is neither the size of the sheet, $\linf$, nor the size of the ruck, $l$, but rather the elasto-gravity length, $\lgr$. 

\subsection{Analytical solution}

The governing equation \eqref{beam_nondim} may be solved subject to the boundary conditions \eqref{sym_BC}--\eqref{contact_BC} to give the vertical displacement of the sheet in the non-contact region
\begin{equation}
W(X) = \frac{1}{\sigma^2} \left[ 1 -  \frac{\cos (\sqrt{\sigma} X) }{\cos (\sqrt{\sigma} L/2)} \right] + \frac{(L/2)^2 - X^2}{2 \sigma}.
\label{lin_sol}
\end{equation} Here, the dimensionless compressive force $\sigma$ must satisfy
\beq
L \sqrt{\sigma}/2 = \alpha
\label{eqn:stress}
\eeq where $\alpha \approx 4.49341$ is the first root of $\tan x=x$. Eqn \eqref{eqn:stress} is the dimensionless version of \eqref{classical_buckling} while \eqref{lin_sol} is the classic shape of a horizontal heavy elastica, and has been derived previously \cite{Vella2009,Wang1986}. 

A relationship between the (unknown) compressive force, $\sigma$, and the imposed end-shortening, $\dL$, is obtained by substituting \eqref{lin_sol} into \eqref{strain_stretch} and yields
\beq
\Delta L = \frac{4 \alpha^2 S \Linf}{L^2} + \frac{5}{768 \alpha^4} L^7.
\label{ruck_eq_soln}\end{equation}  Note that in the absence of extensibility, $S=0$, we recover the result \cite{Vella2009} that
\beq
L=\left(\frac{768\alpha^4}{5}\right)^{1/7}\dL^{1/7}\approx4.844\dL^{1/7},
\eeq which is the precise version of the scaling given in \eqref{eqn:dl_inextens}.

Equation \eqref{ruck_eq_soln} is the central result of our analysis. A sketch of $\dL(L)$ given by \eqref{ruck_eq_soln} with non-zero stretchability $S>0$ (see figure \ref{bifurcation}a) shows that for sufficiently large values of $\dL$, two values of $L$ may give a rucked solution; for small $\dL$ no such solution exists. Thus there is a minimum value of $\dL$ for which rucked states can exist. In particular, this turning point occurs at
\begin{equation}
\Lcrit = \left( \frac{2^{11}\cdot3}{5\cdot7}\right)^{1/9} \alpha^{2/3} (S\Linf)^{1/9}, \quad \dLmin = \left(\frac{3^{16}\cdot5^{2}}{2^{4}\cdot7^{7}}\right)^{1/9} \alpha^{2/3} (S\Linf)^{7/9}. 
\end{equation}
Rucked solutions exist only when the displacement $\dL>\dLmin$ and, at this critical point, the ruck solution has width $\Lcrit$. The critical buckling load is given by substituting this value of $\Lcrit$ into \eqref{eqn:stress}; crucially this load is finite.

The existence of an arbitrarily large buckling load for a ruck was based on the assumption that rucked solutions exist for arbitrarily small end-shortenings, as discussed around \eqref{eqn:infiniteload}. The existence of a minimum end-shortening $\dLmin$ for rucked solutions resolves this paradox but leaves the question of precisely when a ruck should form unanswered: while a ruck is possible for $\dL>\dLmin$ does a system necessarily attain that state as soon as $\dL>\dLmin$? Furthermore, since two rucked solutions are possible for $\dL>\dLmin$, which of these will be observed in practice?

\subsection{Multiple states}

\begin{figure}
\centering
\includegraphics[width=12cm]{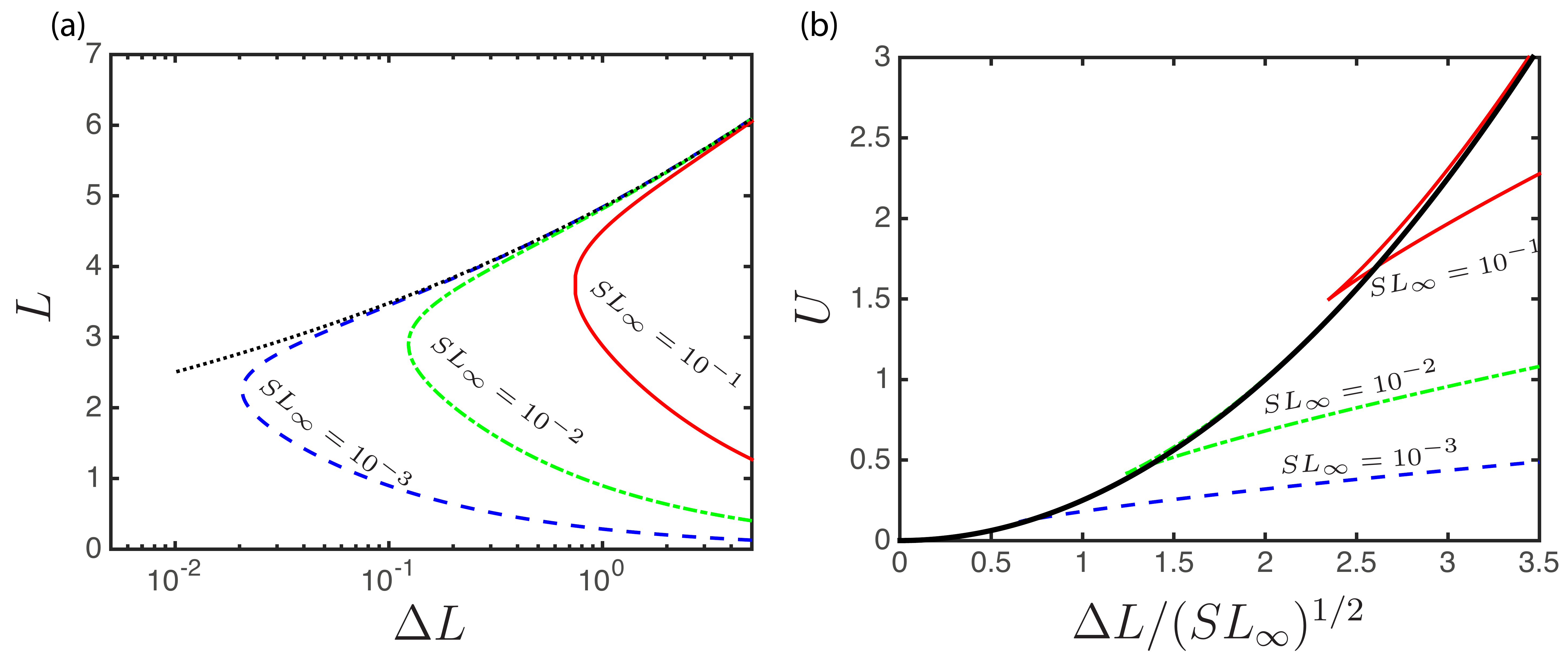}
\caption{(a) The response diagram of the system as a function of the end-shortening, $\Delta L$. We see that ruck solutions become possible only $\Delta L > \Delta L_{\mathrm{min}}(S)$. Results are shown for different values of the parameter $S\Linf$, as indicated, and the relationship for $S=0$, $L\approx4.844\dL^{1/7}$ is also shown (black dotted curve). (b) A comparison of the energy of the buckled solutions (colored curves) and the purely compressed solution (solid black curve) as the end-shortening $\dL$ increases. We see that the buckled solution is not necessarily a global energy minimizer as soon as it appears, but becomes so very shortly after its appearance. Here results are shown for a variety of values of $S\Linf$.  Note that there are two buckled solutions possible for $\dL>\dL_{\mathrm{min}}$ (see a); that with the larger value of $L$ has the lower energy, and is the one that we  expect to see experimentally.  }
\label{bifurcation}
\end{figure}

We have seen that for $\dL<\dLmin$ no rucked solution exists and thus conclude that the the imposed deformation must be accommodated by compression.  However, when $\dL>\dLmin$, there are three possible states that the sheet can adopt: two rucked states (with different values of $L$, see figure \ref{bifurcation}a) and one compressed. 

To determine whether spontaneous formation of a ruck is energetically favourable in comparison to the compressed state, we consider the mechanical energy of the system. The total energy of the system consists of bending, stretching and gravitational energy and may be written
\beq
\frac{E}{2}=\int_0^{\linf/2}\left[\tfrac{1}{2}Bw_{xx}^2+\tfrac{1}{2}(-T)\left(u_x+\tfrac{1}{2}w_x^2\right)+\rhos g t w\right]~\upd x.
\label{eqn:CorrectEnergy}
\eeq Here, the first term in the integral represents bending energy, the second the elastic strain energy and the third the gravitational potential energy. (In \ref{sec:appendix} we show that minimizing this energy subject to an imposed end--end displacement leads to the heavy elastica equation \eqref{beam}.)

Note that when there is no vertical deflection, the energy reduces to
\beq
\frac{U_{\mathrm{flat}}}{2}=\tfrac{1}{4}T\dl=\tfrac{1}{4}\frac{Et}{1-\nu^2}\frac{\dl^2}{\linf},
\eeq or, in dimensionless terms, 
\beq
\frac{U_{\mathrm{flat}}}{2B/\lgr}=\tfrac{1}{4}\frac{\dL^2}{S\Linf}.
\eeq

When considering the energy of the buckled state, it is important to note  that the stretching energy is reduced (compared to the flat case) by a reduction in the compressive force. The contribution of the whole energy that comes from stretching is $\int_0^{\linf/2}(-T)e_{xx}/2~\upd x=(1-\nu^2)T^2\linf/(4 Et)$. In dimensionless terms the energy of the buckled state becomes
\begin{align}
\frac{U_{\mathrm{buckled}}}{2B/\lgr}&=\frac{S\Linf\sigma^2}{4}+\int_0^{L/2}\left(\tfrac{1}{2}W_{XX}^2+ W\right)~\upd X\nonumber\\
&=\frac{4\alpha^4S\Linf}{L^4}+\frac{7L^5}{384\alpha^2}.
\label{eqn:CorrectEnergyNonDim}
\end{align}

The behaviour of the energy of the buckled solutions as the end-shortening increases is shown in figure \ref{bifurcation}b. We see that the buckled solution with the largest value of $L$ has the lower energy of the two ruck solutions (as might be expected since it is the least curved). However, we also note that immediately after the buckled solution appears, it has a slightly higher energy than the unbuckled state and thus is only metastable. (We note that a cusped energy curve with a small portion of the buckled state that is metastable was also observed in a related study \cite{Napoli2015}.) Nevertheless, the length of this metastable branch decreases for decreasing $S L_{\infty}$, and as $\dL$ increases, the energy of this buckled state is considerably lower than that of the unbuckled, purely compressed state. Therefore,  our model predicts that for stiff systems where the stretchability is a regularizing parameter ($S L_{\infty} \ll 1$), the transition from flat to buckled solutions will occur close to $\dLmin$ and without hysteresis; this prediction will be validated in \S \ref{experimental_section}.

\subsection{Validity of approximations}

Our theoretical results have been predicated on the assumption that the deflection of the sheet are small and that, in particular, the slope of the sheet remains small. We have then shown that buckling only becomes possible for $\dL>\dLmin\sim (S\Linf)^{7/9}$. Now, the height of this buckled shape $d\sim L^4\sim \dL^{4/7}$ while the width $L\sim \dL^{1/7}$; hence the typical slope of the delaminated shape is $d/L\sim \dL^{3/7}$. For this slope to be small at the onset of buckling (and hence for our approach to be valid) we therefore need that $\dLmin^{3/7}\ll1$, i.e.~$S\Linf\ll1$. We also need to ensure that the width of the `blister' at onset is smaller than the total length of the sheet, which requires $(S\Linf)^{1/9}\lesssim \Linf$. However, we envisage that this will hold whenever $S\Linf$ is sufficiently small and $\Linf\gtrsim1$.

\section{Experiments}
\label{experimental_section}

To test the predictions of our model, we performed experiments with sheets with a range of thicknesses and Young's moduli. The sheets were made from Polyvinylsiloxane (PVS, supplied by Zhermack) of different stiffness; $E$ was measured directly to be in the range $95\mathrm{~kPa}\leq E\leq 440 \mathrm{~kPa}$. The PVS sheets are made by casting the polymer melt in a rectangular tray and allowing it to cure. The width was not varied systematically but all results presented here had $10\mathrm{~cm}\leq b\leq 14\mathrm{~cm}$ while the length was varied in the range $23.3\mathrm{~cm}\leq\linf\leq29.3\mathrm{~cm}$. By varying the volume of polymer melt used, the thickness of the resulting sheet can easily be varied, but in each case is extremely uniform (to within $4\%$), as has been observed previously \cite{Nasto2013}; here we use sheets with thickness lying in the range $4.8\pm0.2\mathrm{~mm}\leq t\leq 7.7\pm0.2 \mathrm{~mm}$. Detailed values of the experimental parameters, and the values of $S$ to which they correspond, are given in table \ref{table:pars}.

\begin{table}
\centering
    \begin{tabular}{c|c|c | c | c|c}
    $E\mathrm{~(kPa)}$&  $t\mathrm{~(mm)}$ & $\rhos\mathrm{~(kg/m^3)}$ & $\lgr\mathrm{~(cm)}$& $S$ & Symbol   \\ \hline
$440$ & $4.8$ & $1220$ & $4.55$ & $9.3\times10^{-4}$ & $\blacktriangle$, $\triangledown$\\
$180$ & $5.0$ & $1220$ & $3.47$ & $1.7\times10^{-3}$ & $\square$, $\blacksquare$, $\star$\\
$94$ & $7.0$ & $1180$ & $3.51$ & $3.2\times10^{-3}$ & $\Circle$, $\CIRCLE$\\
    \end{tabular}
\caption{Material properties for the three sets of experiments performed here, together with the symbol used to indicate that series of experiments in figure \ref{fig:experiments}.}
\label{table:pars}
\end{table}

The polymer sheets formed in this way can be made to be very soft and relatively thick. However, they are in general tacky, adhering to surfaces. They therefore do not slip easily past the substrate as has been assumed in our model. To overcome this problem, we adhered thin bands of acetate perpendicular to the direction of compression (as can be seen in figure \ref{fig:schematic}b) allowing the sheet to slide smoothly over the substrate without changing its elastic properties noticeably. 

\begin{figure}
\centering
\includegraphics[width=12cm]{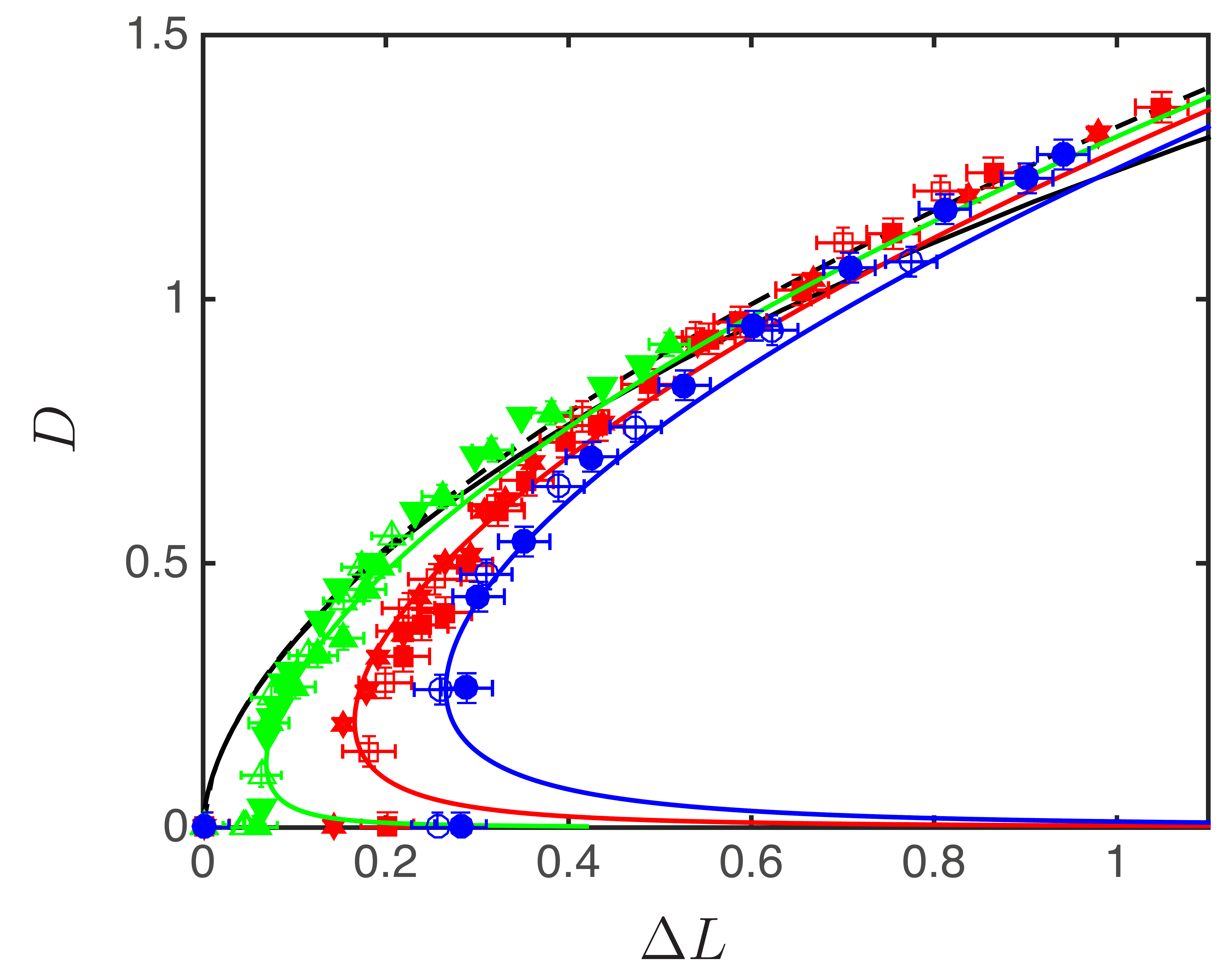}
\caption{Experimental measurements of the dimensionless ruck height, $D=d/\lgr$, with the (dimensionless) end-end compression $\dL$. Results are shown for increasing compression (closed symbols) and decreasing compression (open symbols) with a variety of different stretching stiffnesses: $S=0.00093$ (triangles), $S=0.0017$ (squares and stars) and $S=0.0032$ (circles). The corresponding theoretical predictions are shown by the solid curves. We also plot here the theoretical predictions in the limit of $S=0$: in this limit the linear theory predicts that $D\approx1.326\dL^{4/7}$ \cite{Vella2009} (dashed black curve); the  prediction from the geometrically nonlinear theory \cite{Vella2009} is also shown (solid black curve).}
\label{fig:experiments}
\end{figure}

Figure \ref{fig:experiments} shows the results of experimental measurements of the dimensionless ruck height $D=d/\lgr$ as a function of the end-end compression $\dL$. The ruck height is easier to measure experimentally than the region over which contact is lost, $L$; eqn \eqref{lin_sol} gives that
$$D=\frac{L^4}{32\alpha^4}\left(\alpha^2+2-2\sec\alpha\right)\approx0.0024 L^4.$$
These experiments were performed with three different values of the stretchability parameter $S$ (as described in table \ref{table:pars}) and agree well with the predictions of the model presented earlier for the appropriate values of $S$.

As expected on energetic grounds, the sheets jump to the ruck solution almost as soon as it becomes available and, further,  it is the  solution branch with larger $L$ that is selected. We also note that we do not observe hysteresis in this system: the results when the compression is gradually released (open symbols) collapse with those obtained as the compression is increased (closed symbols), to within experimental tolerances. This distinguishes this problem from apparently related problems such as the delamination of a stiff sheet from a soft foundation \cite{Vella2009PNAS}, where hysteresis is large enough to be easily observed.

\section{Conclusion}

We have studied the onset, or birth, of a ruck in a rug and shown that the finite extensibility of the sheet means that, in reality, rucking can occur at finite end-end compression, and with finite compressive load. This resolves an issue with the standard, linearized but inextensible theory that has caused some confusion in the literature.

Our analysis shows that rather than taking any given size, blisters have a preferred size, below which it is energetically preferrable to accommodate the imposed end--end displacement by compressing the material uniformly. In the delamination of a stiff sheet from a soft foundation, the existence of such a preferred size leads to the formation of a series of blisters as the end--end compression increases \cite{Vella2009PNAS}. One might then naturally ask why does only one ruck form, rather than a whole series, as $\dl$ increases beyond the onset of rucking? To answer this, we note that the energy of a ruck $\sim Bd^2/l^3\sim \dl^{5/7}$ --- dividing a given end--end compression between $N>1$ rucks leads to a total energy $\sim N\times(\dl/N)^{5/7}\sim N^{2/7}$, which is energetically unfavorable compared with a single ruck.

We note that in other problems where non-zero extensibility has been found to play a role, the relevant stretchability parameter has been determined purely from the geometry of the system. For example, the growth of a rod confined within a circular tube is controlled by the ratio of the rod thickness to the cylinder radius \cite{Napoli2015}, while the buckling, snap-through and `ringing' oscillation of beams and arches depends on the ratio of thickness to beam length \cite{Neukirch2012,Pandey2014}. In contrast, for the birth of a ruck in a rug, the relevant stretchability is $S=t^2/(12\lgr^2)$. Rather than being purely geometrical, this $S$ contains the mechanics of the problem via the elasto-gravity length scale $\lgr$. This is a consequence of the importance of two force scales in the problem (gravity and elasticity) and, while not generic, such balances may give rise to similar behaviour when extensibility is included in related systems \cite{Wagner2011,Oshri2015}.

Finally, we note that there are likely other ways in which the divergence of compressive stress might be regularized in this, and other, problems. For example, we expect that a finite substrate compliance might lead to wrinkling of the combined system \cite{Stafford2004,Brau2011} at small end--end compressions but become energetically unfavourable compared to rucking at larger compressions.

 \section*{Acknowledgments}
  
\noindent We are grateful to Alain Goriely and Thomas Lessinnes for discussions, as well as two referees who made numerous helpful suggestions. This work is partly funded by the European Research Council (ERC) under the European Union's Horizon 2020 research and innovation programme (grant agreement No 637334).

\appendix

\section{Energy and a variational principle\label{sec:appendix}}
Considering only half of the system, i.e.~$0\leq x\leq \linf/2$, the functional describing the system consists  of the bending, stretching and gravitational energies of the beam, together with a constraint associated with the end--end compression. This constraint is enforced by a Lagrange multiplier  $F$ so that the appropriate functional is
\beq
{\cal U}=\int_0^{\linf/2}\left[\tfrac{1}{2}Bw_{xx}^2+\tfrac{1}{2}(-T)\left(u_x+\tfrac{1}{2}w_x^2\right)+\rhos g t w\right]~\upd x-F\left[\tfrac{\dl}{2}+u(\linf/2)\right].
\label{eqn:TotalEnergy}
\eeq Here, the first term in the integral represents bending energy, the second the elastic strain energy and the third the gravitational potential energy. 

On first sight, it appears that the factor $\tfrac{1}{4}Tw_x^2$ within the integral means that the coefficient of $Tw_{xx}$ in the corresponding Euler--Lagrange equations will be $1/2$, rather than unity in \eqref{beam}. However, a more careful consideration reveals that the heavy elastica equation  \eqref{beam} \emph{is}, in fact, recovered. To see this, we first eliminate $T$ from \eqref{eqn:TotalEnergy}, using Hooke's law \eqref{hooke}, and write $u(\linf/2)=\int_0^{\linf/2}u_x~\upd x$. Performing the variation with respect to $u$ we find that
\beq
F=\frac{Et}{1-\nu^2}\left(u_x+\tfrac{1}{2}w_x^2\right)=-T.
\eeq  At the same time, the variation with respect to $w$ gives
\beq
Bw_{xxxx}+Tw_{xx}+\rhos gt=0,
\eeq which is the expected heavy elastica equation, eqn \eqref{beam}.

From this, we see that \eqref{eqn:CorrectEnergy}  gives the energy of the system and is consistent with the natural variational principle used to derive the heavy elastica equation \eqref{beam}.

\end{document}